\documentclass[onecolumn,showpacs,preprintnumbers,amsmath,amssymb]{revtex4}

\usepackage{graphicx}
\usepackage{dcolumn}
\usepackage{bm}

\def\be{\begin{equation}}
\def\ee{\end{equation}}

\begin{document}

\title{A Brane model with two asymptotic regions.}

\author{Musongela Lubo}
 \affiliation{The Abdus Salam International Centre for Theoretical Physics I.C.T.P\\
	P.O.Box
586\\
34100 Trieste, Italy.\\
}
\email{muso@ictp.trieste.it}

\date{\today}
\begin{abstract}
Some brane  models rely on a generalization of the Melvin
magnetic universe including a complex scalar field among the  
sources. We argue that 
the geometric  interpretation of Kip.S.Thorne of 
this geometry restricts the kind of potential a complex  scalar 
field can display to keep the same asymptotic behavior. 
While a finite energy is not obtained for 
a Mexican hat potential in 
this 
interpretation, this is the case for a potential displaying a broken phase 
and an unbroken one. We use for technical simplicity and illustrative 
purposes an ad hoc potential which however shares some features with those
obtained in some supergravity models. We construct a sixth dimensional
cylindrically symmetric solution which has two asymptotic regions: the
Melvin-like metric on one side and a flat space displaying a conical singularity on 
the other. The causal structure  of the configuration is discussed.
Unfortunately, gravity is not localized on the brane.
\end{abstract}

\maketitle

\section{Introduction}
\par Among the most important characteristics of cosmic strings is the
existence of a symmetry axis and the concentration of energy
around this axis \cite{LABEL1}. Taking  gravity into account, the existence of 
a symmetry axis implies
cylindrical symmetry for the metric as well. The static
cylindrically symmetric solutions of Einstein equations in vacuum
in $4D$ are of the Kasner type: they are parameterized by three constants
obeying two constraints. The vanishing of the energy-momentum
tensor in the asymptotic region  implies that the geometry must approach a Kasner line
there. As the energy momentum tensor corresponding to these axial
configurations implies the invariance of the metric under boosts,
one is left with only two Kasner geometries : a flat space
presenting a conical singularity and a Melvin like space
\cite{LABEL2}. The first case leads to the well known cosmic
strings. The second solution, written in a particular system of coordinates,
displays circles of  decreasing circumferences for increasing ``radii''$\rho$.
This feature has been analyzed by Kip.S.Thorne \cite{thor1,thor2}. The interpretation is 
that   $\rho=\infty$ is the point at infinity on the symmetry axis.

The Melvin solution has high dimensional generalizations which can be
used to build brane models \cite{will1,will2}. In recent works, complex 
scalar fields have been 
incorporated into the picture \cite{cigar}. In this article we address the same question.
However, we use the Kip S.Thorne interpretation to fix the boundary conditions;
the coordinate $\rho$ being the point at infinity on the symmetry axis, the 
angular coordinate is not well defined there, just as for the polar coordinates
at the origin on the 
plane. A cylindrically symmetric complex field must thus vanish at that point.
We obtain that no static cylindrical solution can be obtained with the usual
Higgs potential. We then exhibit a toy model for which this can be achieved
and discuss its characteristics.

This paper is organized as follows. In the second section we review the
geometric interpretation of The  Melvin magnetic solution in four dimensions.
The third section incorporates the scalar field, set the boundary conditions and
displays the numerical solution obtained. First, an Abelian-Higgs Lagrangian 
is coupled to
the Einstein-Hilbert one. Looking for an axially symmetric
configuration which displays the second special Kasner line element far from the source
, the preceding section
imposes the vanishing of the vector and the scalar field as the
coordinate $\rho$ goes to
infinity. This  results in
a divergent inert mass per unit length if the v.e.v of the Higgs
field does not vanish. If it does, one simply recovers the Melvin solution. 
We then construct, for illustrative purposes, a 
 potential for which the behavior of the scalar is non trivial. This potential
has two minima, one of them being at zero. This has similarities with some
potentials obtained in some supergravity \cite{cigar} models and some non commutative 
models \cite{nonco}. The constructed solution 
interpolates between them. The minimum corresponding to a vanishing value of
the field is attained in the Kasner-like asymptotic region while the non 
vanishing value corresponds to a flat space presenting an angular deficit.
The classical trajectories of
neutral particles in this geometry are analyzed in section four. We show
that the trajectories of massive particles in this geometry are 
bounded.
We also study massless particle trajectories
and discuss the causal structure of the solution.
An appendix is devoted to the way the numerical approximation has
been computed.

The coupling of scalar fields to gravity leads to many classical solutions
\cite{di1,di2,di3,di4,di5,di6,di7,di8}.
The introduction of a scalar field among the sources leading to a Melvin type
universe has been considered before \cite{maeda}. There are three main
differences between our work and the previous papers. Firstly, the scalar field
considered here is not a dilaton so that its coupling to gravity is ordinary.
The involved
potentials are not the same.
Secondly, the scalar field here is complex, contrary to \cite{maeda} where
it is real. The vanishing of the scalar field on the symmetry axis is not
required when it is  real. On the contrary, this becomes mandatory when it is
complex, just like for the Higgs field on the cosmic string core.  But, the more
important difference is that here we have a configuration with two asymptotic
regions. Like in \cite{gorba}, we do not have a delta function for  the brane.

\section{The Melvin solution in $4D$.}

In this section we  review the geometric interpretation of the Melvin solution.
To keep the discussion as simple as possible, we actually analyze its
asymptotic limit in four dimensions. The conclusions are however the same
in the presence of extra dimensions and the addition of matter.

\par We will study a system consisting of a self gravitating Maxwell system
in $d$ dimensions. The classical field equations are derived from
the action
\be
\label{action}
S=\int d^dx \sqrt{\vert g \vert}\left[ - \frac{1}{4} F_{a b} F^{a b}
+ \frac{R}{16\pi G} \right].
\ee
The solution which generalizes the Melvin universe in $d$ dimensions is 
given by the following expressions of the metric and the Maxwell tensor:  
\begin{eqnarray}
\label{louko}
ds^2 &=& \left( 1 + \frac{\rho^2}{a^2} \right)^{2/(d-3)} \left( \eta_{\mu \nu} 
dx^\mu dx^\nu - d\rho^2 \right) - \rho^2
 \left( 1 + \frac{\rho^2}{a^2} \right)^{-2}  d\phi^2    \quad , \nonumber\\
F &=&  
B_0 \left( 1 + \frac{\rho^2}{a^2} \right)^{-2} \rho \,  d\rho \wedge d\phi  \quad  ,    
\end{eqnarray}
$a$ and $B_0$ being  dependent constants.
This solution has been used in the brane world scenarios. One of its important 
properties is that  the brane can have positive tension and the closure 
of the bulk provides a singularity-free boundary condition for solutions
that contain black holes and gravitational waves on the brane
\cite{will1,will2}.

The characteristic of this metric on which we will put the emphasis
is the fact that the circumference of  circles obtained by letting only
$\phi$ non fixed tend to zero as the coordinate $\rho$ goes to infinity.
The Lagrangian from which the solution given in Eq.(\ref{louko}) follows
did not contain any scalar field. The question we address in this paper
is: if one adds a scalar field to the picture, which kind of  
potential allows the same behavior for the metric? We will argue that a 
geometric interpretation of this  property of the metric
gives an important restriction. 

The features observed by Kip.S.Thorne for the Melvin  solution 
\cite{mel1,mel2}
are already present in a vacuum solution: the Kasner line element \cite{LABEL2}
 which is 
obtained by taking $a=0$ in Eq.(\ref{louko}). 
The  simplest example where this behavior can be analyzed is  the sphere.
Introducing the polar stereographic coordinates ($ r, \theta$), the 
metric of a sphere of radius $\sigma$ reads \cite{geo}
 \be \label{4}
ds^2 = {16\sigma^4\over {(4\sigma^2+ r^2)^2}} dr^2 + {16 \sigma^4
r^2\over {(4\sigma^2+ r^2)^2}} d\theta^2 \quad . \ee 
For large values of the coordinate $r$, the coefficient
$g_{\theta\theta}$ becomes a decreasing function . If one interprets $r$ 
  as the radius, a circle of infinite
radius turns out to be of null length. This is obvious
since $r = \infty$ corresponds to the point at infinity on the
plane which is mapped into the north pole by the  stereographic
projection; $ r = \infty $ is just the north pole. When the
coordinate $ r$ vanishes, one has another circle displaying a
vanishing circumference : the south pole. The two are on the
symmetry axis. Introducing the variable 
\be \label{5} r_{\ast} =
2\sigma \arctan (r/2\sigma) \quad , \quad
{\rm  the \quad metric \quad reads} \quad
 ds^2 = dr^2_{\ast} + \sigma^2 \sin^2
\left({r_{\ast}\over {\sigma}} \right) d\theta^2 \quad .
 \ee The relation
between $ r_{\ast}$ and $r$ is bijective provided that $r_{\ast}
\in [ 0, \pi \sigma ]$. The points located on the symmetry axis
once again are those for which the coefficient $g_{\theta\theta}$
vanishes.

\section{The $6D$ extension with a complex scalar field.}

Before proceeding, let us point out that the tangential
Maxwell vector field in the Melvin 
solution vanishes when $\rho$ goes to infinity, in accord with the geometric
interpretation. We now wish to include a scalar field in the picture, in the 
presence of extra dimensions. We choose a sixth dimensional model essentially
because in this case one can naturally obtain chiral fermions.

Let us first consider a scalar displaying a Higgs potential; the
matter action is then
\be
\label{15}
S=\int d^6 x \sqrt{-g} \left[{1\over 2} D_{\mu} \Phi D^{\mu} \Phi^{\ast}-
{\lambda\over 4} (\Phi \Phi^{\ast} - v^2)^2 \right].
\ee
The $U(1)$ charge $e$ is embodied in the covariant derivative
\mbox{$D_{\mu}\Phi = \partial_{\mu}\Phi - i eA_{\mu}\Phi$}. For a static
cylindrically symmetric configuration, the ansatz can be given the
form
\begin{eqnarray}
\label{16}
ds^2 &=& \beta^2(\rho) \eta_{\mu\nu} dx^\mu dx^\nu
- \gamma^2(\rho) d\rho^2 - \alpha^2(\rho) d\phi^2 \quad  {\rm where} \quad
\mu = 0, \cdots 3  \quad , \nonumber\\
\Phi &=& v f(\rho) e^{i\phi} \quad {\rm and} \quad A_{\phi} = 
\frac{1}{e} (1-p(\rho)) \quad .
\end{eqnarray}
The cosmic string solution has been extensively studied in the
literature. In that configuration, the smoothness of the geometry
on the symmetry axis is guaranteed by going to the gauge $\gamma(\rho)=1$ 
and imposing the boundary conditions
\cite{LABEL2} 
\be \label{18} \alpha(0) = 0 , \alpha'(0) = 1 \quad , 
\ee while the matter fields are non singular on the
core provided that \be \label{19} f(0) = 0\ ,\ p(0) = 1  \quad . \ee
The vanishing  of the energy density in the asymptotic region implies
\be \label{20} f(\infty) = 1 \quad , \quad p(\infty) = 0 \quad . \ee 

What happens if we want the metric to display the same asymptotic 
behavior than in Eq.(\ref{louko}) when the coordinate $\rho$ goes to infinity?
In the previous section,
we argued that $ \rho = \infty$ is the point at infinity on the
symmetry axis. To have a regular cylindrically symmetric configuration, the Higgs 
and the tangential
vector field must vanish there:
\be \label{21} f(\infty) = 0\quad
{\rm and} \quad p(\infty) = 1 \quad . \ee
\par Extracting the expression of the integrand of the
inertial mass from Eq.(\ref{15}) one
has 
\be \label{22} 
\epsilon(\rho) =   \sqrt{\vert g \vert}  \left[{1\over 2}
g^{\rho\rho} |D_\rho \Phi|^2 + {1\over {2 }} g^{ \phi \phi} |D_\phi
\Phi|^2
 + {1\over 4} F_{\rho
\phi} F^{\rho \phi} + {\lambda\over 4} (\Phi^{\ast} \Phi- v^2)^2
\right]. \ee In the asymptotic region (i.e. $ \rho \rightarrow
\infty$) one has $ \sqrt{- g}  \sim \rho^{7/3}$; the volume element is
not bounded. The first three terms decrease in the asymptotic
region provided that $f(\rho)$ and $p(\rho)$ approach constants there;
this is already satisfied by Eq.(\ref{21}). The contribution of
the Higgs potential in this part of the space is   reduced to the
integral of  $ \rho^{7/3} \lambda v^4 $. This has a chance to converge
only when $v=0$. Then, the parameterization given in Eq.(\ref{16}) does
not apply; one can however define a dimensionless function
associated to the Higgs field by using the Newton constant. Doing this,
we obtained a
vanishing scalar field for any value of the parameters.

Physically
this can be understood as follows. Forcing the scalar field to go
from zero to zero as $\rho$ goes from  zero to infinity, one obtains
that it vanishes identically since there is no source. Such a
source would be for example a local maximum of the potential but
as the vacuum expectation value vanishes, such a maximum does not
exist. In fact, one recovers the 
Melvin universe.

Is it possible to construct a solution with a non vanishing scalar field? To do this we
need a potential which vanishes with the scalar field so that the
minimum is attained at spatial infinity. We also need a local
maximum which will correspond to a source. These conditions are
for example satisfied by the gauge invariant potential \be
\label{28}
  V( \Phi) = \lambda  e^{w^2 \Phi \Phi^*} \Phi \Phi^{*} (  \Phi \Phi^{*} -  v^2
  )^2 \quad .
  \ee
The maximum is attained
at $ \phi = \pm v \sqrt{\sqrt{2}-1}  $ while there are  three  minima , at  $
 \phi  = 0 , \pm v $. The U(1) symmetry is broken spontaneously in
the last two vacua and preserved in the first one. We disregard the
 renormalizability since we are interested only in classical solutions. This 
 potential, although purely ad hoc, shares with the one appearing in
 \cite{cigar} 
 \begin{equation}
 \label{pot}
 V(\phi)= 2  e^{\phi \bar{\phi} } (\phi \bar{\phi}  )^{p-1}
 \left[ 2 (p+ \phi \bar{\phi} )^2 - 5 \phi \bar{\phi}  \right]
 \end{equation}
 the fact that it is the product of an exponential and a polynomial. The 
 difference is the fact that in Eq.(\ref{pot}), one needs to have $p\geq 1$ to have
 the vanishing value of the field as a vacuum but then there is no other 
 vacuum. In \cite{cigar}, it was argued that the potential of Eq.(\ref{pot}) could be seen
 as inspired from some supergravity model, with a particular choice of the
 Kahler structure. Let us also mention that in models in which one works with
 non commutative spaces such as ${\cal M} \times M_n$ where ${\cal M}$
 denotes the Minkowski space and $M_n$ the set of $n \times n$ matrices, one also
 obtains potentials displaying symmetric  vacua. 
The attitude adopted here is like the one concerning analytical solutions for 
self gravitating domain 
walls \cite{selfgra1,selfgra2,selfgra3}. One knows that a  potential which is a cosine of a scalar 
field achieves the desired goal, although it is not renormalizable. In the 
same way, one introduces ad hoc
potentials for  quintessence \cite{quint1,quint2,quint3}. Some of them 
are negative powers of a scalar field and so lead to non renormalizable
theories. Our only aim is to show that solutions
with finite energy exist for specific potentials. Moreover,
some potentials displaying  symmetric vacua have been used as candidates 
for dark matter \cite{be,wei}.

We now wish to construct a solution which interpolates between two vacua, say
$\vert \Phi \vert =v$ and $\vert \Phi \vert=0$. Our previous discussion tells us that the region where
the field goes to its unbroken phase can not be Melvin-like. As we simply 
require cylindrical symmetry, we can choose that asymptotic region to be like
the far region of a cosmic string. 

The Einstein equations will be written in the form
\begin{equation}
\label{ein}
R_a^b= - 8 \pi G \left( T_a^b - \frac{1}{4} T \right)
\end{equation}
where the factor $1/4$ comes from the dimension of the space time.
In sixth dimensions, asking our action to be a pure number means the 
self coupling $\lambda$ and the gauge coupling $e$  are dimensionfull. We thus 
can write our equations in terms of the dimensionless parameters
\begin{equation}
\label{para}
\mu = G v^2 \quad , \quad \nu = \lambda^2 G^{-3} \quad , \quad \tau = e^2 v
\quad {\rm and} \quad \sigma = w^2 v^2 \quad .
\end{equation}
Note that here the dimension of $v$ is an inverse length square. We will use
the dimensionless
length $x$  given by 
\begin{equation}
\label{len}
\rho = L \, x \quad  \quad {\rm where} \quad L = \frac{1}{2 \sqrt{\pi} 
\mu^{5/4} \nu^{1/4}} \frac{1}{\sqrt{v}} 
\end{equation}
and  the  dimensionless functions 
\begin{equation}
 \alpha(\rho) = \sqrt{\pi} \sqrt{\frac{\mu}{\tau}} 
 \frac{1}{\sqrt{v}} A(x) \quad , \quad \beta(\rho) = B(x)
 \quad , \quad \gamma(\rho) = \bar{\gamma}(x) \quad , \quad
 f(\rho) = F(x) \quad , \quad  p(\rho) = P(x) \quad . \quad
\end{equation} 
The field equations read
 \begin{eqnarray}
 \label{231}
& &  e^{\sigma F^2(x)} B(x) F^2(x) \bar{\gamma}^2(x) (1-F^2(x))^2 +
 \frac{A'(x)B'(x)}{A(x)} + 3 \frac{B'^2(x)}{B(x)} - 
 \frac{B'(x) \bar{\gamma}'(x)}{\bar{\gamma}(x)} -
 2 \frac{B(x) P'(x)^2}{A^2(x)}+ B^{''}(x) = 0 \quad ,  
\end{eqnarray}
\begin{eqnarray} 
\label{232} 
& & \frac{1}{4} e^{\sigma F^2(x)} B(x) F^2(x) \bar{\gamma}^2(x) (1-F^2(x))^2 +
 2 \pi \mu B(x) F'^2(x) - \frac{B(x) A'(x) \bar{\gamma}'(x) }{4 A(x) 
 \bar{\gamma}(x)} - \frac{B'(x) \bar{\gamma}'(x)}{\bar{\gamma}(x)} +
 3 \frac{B(x) P'^2(x)}{2 A^2(x)} \nonumber\\
& & + \frac{B(x) A''(x)}{4 A(x)} 
 +  B''(x) = 0  \quad , 
\end{eqnarray}
\begin{eqnarray}
\label{233}
& & e^{\sigma F^2(x)} B(x) F^2(x) \bar{\gamma}^2(x) (1-F^2(x))^2 +
\frac{2 \tau}{\pi \mu^{5/2} \sqrt{\nu}} 
\frac{F^2(x) \bar{\gamma}^2(x) P^2(x) }{A(x)} + 4 \frac{A'(x) B'(x)}{B(x)} -
\frac{A'(x) \bar{\gamma}'(x)}{\bar{\gamma}(x)} \nonumber\\
& & + 6 \frac{P'^2(x)}{A(x)} + A''(x) =0 \quad , 
\end{eqnarray}
\begin{eqnarray}
\label{234}
& & \frac{\tau}{2 \pi \mu^{5/2} \sqrt{\nu}} 
F^2(x) \bar{\gamma}^2(x) P(x) - \frac{A'(x) P'(x)}{A(x)} + 
4 \frac{B'(x) P'(x)}{B(x)} - \frac{\bar{\gamma}'(x) P'(x)}{\bar{\gamma}(x)}
+ P''(x) = 0 \quad , 
\end{eqnarray}
\begin{eqnarray}
\label{235}
& & \frac{1}{2 \pi \mu} e^{\sigma F^2(x)} F(x) \bar{\gamma}^2(x)
( 1+(-4+\sigma) F^2(x)+(3-2 \sigma) F^4(x) + \sigma F^6(x)) - 
\frac{\tau}{4 \pi^2 \mu^{7/2} \sqrt{\nu}} 
\frac{F(x) \bar{\gamma}^2(x) P(x)}{A^2(x)} \nonumber\\
& & \frac{A'(x) F'(x)}{A(x)}+ 4 \frac{B'(x) F'(x)}{B(x)}
- \frac{F'(x) \bar{\gamma}'(x)}{\bar{\gamma}(x)} + F''(x) = 0 \quad .
\end{eqnarray}

We now relax the assumption that the coordinate $\rho$ goes from zero to 
infinity but rather take it to go from minus to plus infinity. This is like
parameterizing  every point of the sphere not by $\theta,\phi$ but by 
$\phi$ and
its distance from the equator on a meridian. The upper hemisphere
would have positive $\rho$ while  the lower  would correspond to
negative values of that length coordinate. In our case, the region where
$\rho \rightarrow -\infty$ will correspond to a cosmic string-like geometry
while $\rho \rightarrow \infty$ will be associated with the Melvin-like
behavior.

So, the boundary conditions are that when $\rho \rightarrow - \infty$, one is in the far region of the
cosmic string solution:
\begin{equation}
\label{far1}
A(x) \sim  x \quad ,  \quad 
B(x) \sim 1  \quad , \quad \bar{\gamma} \sim 1 
\end{equation}
while for $\rho \rightarrow  \infty$
one enters the asymptotic region of the Melvin solution:
\begin{equation}
\label{far2}
A(x) \sim  \frac{a^2}{x} \quad ,   \quad
B(x) \sim \bar{\gamma}(x) \sim \left( \frac{x^2}{a^2} \right)^{1/3} \quad .
\end{equation}

To give an  illustration, we need to solve the above set of coupled non linear
differential equations. This has been done for the simplest choice of the
parameters: $\mu=\nu=\tau=\sigma=1$; the details concerning the numerical treatment
 are given 
in the appendix. The behavior of the different fields has been given in the 
pictures FIG.1, FiG.2, and FiG.3. Roughly speaking, the region $\rho=0$ is where the transition between
the two regions takes place; it is also the place where the energy is
concentrated.

The trapping of gravity is ensured when the condition
\begin{equation}
\int dx^4 dx^5 g^{00} \sqrt{\vert g \vert} < \infty \quad 
\end{equation}
is satisfied.
From the asymptotic behavior of the metric, one sees this is not true
for the solution we constructed. Let also remark that a change of coordinate can now be made
so that for example $\bar{\gamma}=1$.

\begin{figure}
\includegraphics{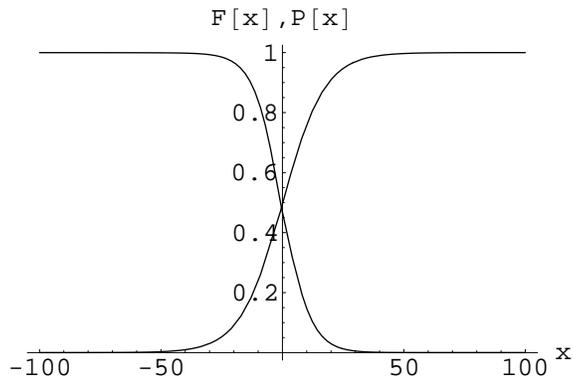}
\caption{The scalar and vector fields are plotted in terms of the coordinate $x$.
 }
\end{figure} 
\begin{figure}
\includegraphics{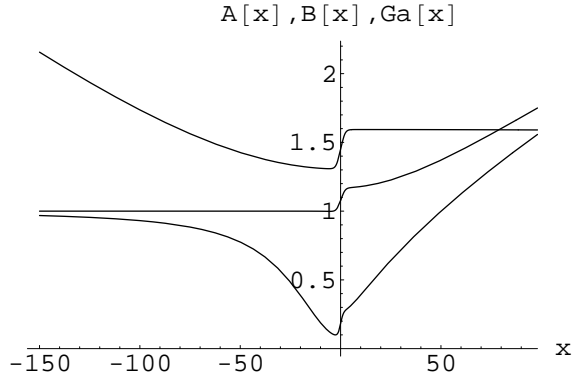}
\caption{From top to bottom, the components $A(x), B(x), \bar{\gamma}(x)$ of the metric  
}
\end{figure}
\begin{figure}
\includegraphics{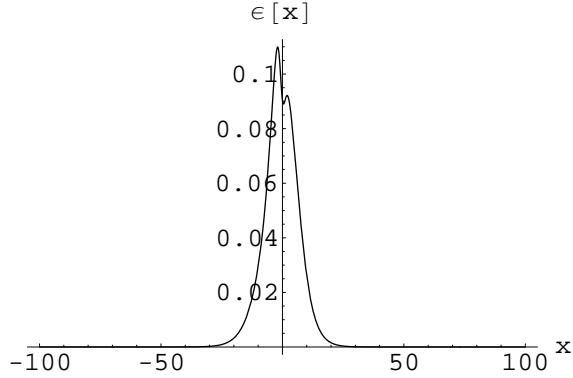}
\caption{The energy density $\epsilon(x)$}
\end{figure}

\section{ The classical trajectories.}

Let us first consider the case of  massive particles. The
metric specified in Eq.(\ref{16}) has  cyclic coordinates ;
its geodesics are by way of consequence characterized by 
constants of motion. Introducing the
proper time per unit mass $\tau$, the energy-momentum relation
reads 
\be \label{31} \left( {d\rho\over{d\tau}} \right)^2 =
\left( -1 - \frac{k_1^2}{\alpha^2(\rho)} + \frac{k_2}{\beta^2(\rho)}  \right)
\frac{1}{\gamma^2(\rho)}  \quad .
\ee 
A physical motion is characterized by a real velocity. The asymptotic
behavior of the metric displayed in Eq.(\ref{far1},\ref{far2}) shows that the
trajectory of a massive particle never attains the point at infinity on the 
Melvin branch; however it has access to the region at infinity on the
string branch.

\par  The causal structure is found by analyzing particular null geodesics.
The bounded null coordinates in the new background are given by
$\bar u = c^{st}$ or $\bar v = c^{st}$ with \be \label{33}
\left.\begin{array}{l}
\bar u \\
\bar v 
\end{array}\right\rbrace = \arctan [t \mp \sigma(\rho)]
  \quad ,
\ee
  where
\be
\label{34}
   \sigma(\rho) = \int_0^\rho d\xi \frac{\gamma(\xi)}{ \beta(\xi)} \quad .
\ee
In FIG.4, we have drawn the Penrose-Carter diagram of the solution.

\par The embedding of the two dimensional metric containing only $t$ 
and $\rho$ in the  Euclidean space can be
realized by the surface of revolution
\be
\label{36} Z(r) = \int_0^r  dy  \sqrt{c^2(y) - 1} \quad , \quad
c(y) = \gamma(\alpha^{-1}(y)) \frac{d}{dy} (\alpha^{-1}(y) ) \quad .
\ee
The limiting behavior of the metric shows that in the region 
$\rho\rightarrow - \infty$,
\begin{equation}
Z(r) \sim c^{st} r \quad ;
\end{equation} 
no restriction is imposed on $r$ and the surface is a cone.
On the contrary, as
$\rho\rightarrow \infty$,
\begin{equation}
Z(r) \sim  \int_{r_\star}^r dy \left[ \left(\frac{a}{y}\right)^{16/3} - 1  \right]^{1/2} \quad ;
\end{equation} 
there is a maximal circumference.

\begin{figure}
\includegraphics{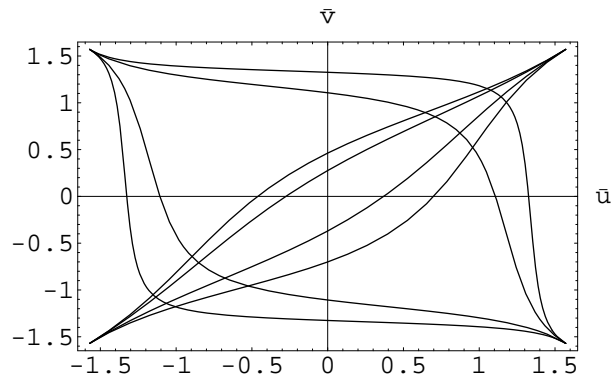}
\caption{The causal structure of the
  solution is given. The time-like
     infinities are $ I^{+}( \bar{u} = \bar{v} = \pi/2 ) $ and $ I^{-} ( \bar{u} = \bar{v}
      = - \pi/2 ) $. As $ r$ can change sign, there are two space
      infinities : $ I^{>}_{o}( - \pi/2 , \pi/2 ) $ and  $ I^{<}_{o}(  \pi/2 , - \pi/2) $.
The curves which begin at $ I^{-} $ and end at $  I^{+}$ correspond to
 fixed values of $r$ while the others correspond to fixed $t$. }
\end{figure}

\section{ Conclusions. }
\par 

We have constructed a sixth dimensional  cylindrically symmetric self gravitating 
configuration. It has two asymptotic regions: one corresponding to a flat
space time with a deficit angle and the other to the second special Kasner line
element. The part of the geometry which displays a deficit angle
can be realized as a cone with a smoothed apex. The second special Kasner 
geometry, on the other side, can be seen as a tube with a  decreasing radius.
Gluing the two, one obtains something close to a funnel. The causal structure 
of this geometry was studied. It should be stressed that such a configuration,
with two topologically different boundary regions, is not possible with static
spherically symmetric
configurations; the Birkhoff theorem forbids this.

The  potential considered is non renormalizable but our discussion shows that 
it is one of the simplest which allows boundary conditions compatible with the
geometric interpretation of the second special Kasner geometry.

Like in \cite{shap1}, we have built a configuration which has two
different asymptotic regions. In our case we have a space with a conical
singularity on one side and a Melvin-like solution on the other, while
in the preceding one
there are two $Ads_5$ space times glued together along a three brane.
The fact that our solution is not asymptotically flat and the non localization
of the four dimensional graviton is similar to \cite{shap2}. Among the 
priorities  which should be addresses if a more realistic model
is built along these lines is the construction of realistic Abelian and
non Abelian four dimensional models \cite{shap3,shap4}.

What we have learned in this work is basically that if one wants to 
include a complex scalar field possessing a winding number on a Melvin-like
solution, one needs a particular kind of potential. Our solution does not 
trap gravity. Nevertheless, the model may still have
some phenomenological interest. Although we have not made here the appropriate
analytic computations, one can not rule out at this stage the possibility of
having a quasi localized four dimensional graviton on the brane like in
\cite{ruba1}. In this model, it was shown that Newton's law of gravity was
valid only between two length scales fixed by the theory.

\underline{Acknowledgments}
We thank  A.D. Dolgov, Shankaranarayanan.S, R.Jeannerot and  I.Dorsner 
for useful criticisms.

\appendix

\section{Numerical Considerations.}

Our numerical approximation relies on a symbolic approximation of the 
fields  which can then be improved by a relaxation method.
The function $F(x)$ goes from $1$ to $0$ when the argument goes from
$-\infty$ to $\infty$. The ansatz is taken
to be
\begin{equation}
F(x)= \frac{1}{2} (1- \tanh{(f_0^2 x)}) 
\frac{f_1^2 +  f_2^2 x^2}{1+ f_2^2 x^2} \quad .
\end{equation}
In the same way, one has
\begin{equation}
P(x)= \frac{1}{2} (1+ \tanh{(p_0^2 x)}) 
\frac{p_1^2 +  p_2^2 x^2}{1+ p_2^2 x^2} \quad .
\end{equation}
The asymptotic behavior of the metric displayed in 
Eqs.(\ref{far1},\ref{far2}) is taken onto account by the following
functions:
\begin{eqnarray}
A(x)&=& \frac{1}{2} (1-\tanh{(a_0^2 x)}) (a_1^2 + a_2^2 x^2)^{1/2} +
      \frac{1}{2} (1+\tanh{(a_3^2 x)}) (a_4^2 + a_5^2 x^2)^{-1/2}
      \quad , \nonumber\\
B(x)&=& \frac{1}{2} (1- \tanh{(b_0^2 x)}) 
\frac{b_1^2 + b_2^2 x^2}{1+ b_2^2 x^2} +
\frac{1}{2} (1+ \tanh{(b_3^2 x)}) 
(b_4^2 + b_5^2 x^2)^{1/3}  \quad , \nonumber\\
\bar{\gamma}(x) &= & \frac{1}{2} (1- \tanh{(g_0^2 x)}) 
\frac{g_1^2 + g_2^2 x^2}{1+ g_2^2 x^2} +
\frac{1}{2} (1+ \tanh{(g_3^2 x)}) 
(g_4^2 + g_5^2 x^2)^{1/3} \quad .
\end{eqnarray}

These parameters are not all independent, due to the fact one constant($a$)
drives the asymptotic behavior of the functions $\alpha(\rho),\beta(\rho),
\gamma(\rho)$ and $P(\rho)$ simultaneously(see Eq.(\ref{louko})). This has been taken into account.
For simplicity,
we introduce $\alpha_5= \sqrt{a_5}$.

For the true solution, all the right members of the equations given in 
Eq.(\ref{231}), $\cdot$ Eq.(\ref{235}) ,
which we
denote $ODE_1(x), \cdots, ODE_5(x)$, vanish. To obtain an initial
approximation for the relaxation method, the idea is to look
 for the values of
the coefficients $f_0,\cdots, g_5$ for which the integral
\begin{equation}
eq(x) = \int_{-\infty}^{\infty} dx \left( ODE_1^2(x) + ODE_2^2(x) + 
ODE_3^2(x) +
ODE_4^2(x) + ODE_5^2(x)  \right)
\end{equation}
is minimal. 

Rather than computing the integral, we approximated the 
surface to which it corresponds by a sum of rectangles. The values of 
the parameters we found are given below. Plotting the functions 
$ODE_k(x)$, one finds an error of the order $10^{-2}$ 
{\em on the entire real axis}.
\begin{eqnarray}
f_0 &=& 0.29525844042277477 \, , \, 
f_1 = 0.9658023897961495 \, , \,  
f_2 = 0.21489954200347053 \, , \,
p_0 = 0.24041269454525904 \, , \nonumber\\
p_1 &=& 0.9931697892207005 \, , \, 
p_2 = 0.4970289653744458  \, , \,
a_0 = 0.8317552649976594 \, , \, 
a_1 = 1.3064942111721376 \, , \nonumber\\
a_2 &=& -0.011436440955002887 \, , \, 
a_3 = 0.8012738041970792 \, , \, 
a_4 = 0.6275485649917462 \, , \, 
\alpha_5 =0.019766866475879406 \, , \nonumber\\
b_0 &=& 0.823008493904919 \, , \,
b_1 = 0.9933291170842984 \, , \, 
b_2 = 0.42807374544606963 \, , \, 
b_3 = 0.7966722551728198 \, , \nonumber\\ 
b_4 &=& 1.2646707186251818 \, , \, 
g_0 = 0.7842562408108704 \, , \, 
g_1 = -0.2977638255573934 \, , \, 
g_2 = -0.0349623024909614 \, ,  \nonumber\\ 
g_3 &=& 0.8326686788126545 \, , \,
g_4 = 0.14292786831126408 \quad .
\end{eqnarray}

\end{document}